\documentclass{andp2012}
\usepackage[english]{babel}

\newtheorem{thm}{Theorem}[section]

\newtheorem{lem}[thm]{Lemma}

\newtheorem{remark}[thm]{Remark}
\numberwithin{thm}{section}

\newcommand{\R}{{\mathord{\mathbb R}}}
\newcommand{\N}{{\mathord{\mathbb N}}}

\newcommand{\C}{{\mathord{\mathbb C}}}

\newcommand{\E}{{\mathord{\mathbb E}}}

\DeclareMathOperator{\tr}{tr}
\DeclareMathOperator{\Tr}{Tr}
\DeclareMathOperator{\Span}{span}

\DeclareMathOperator{\diag}{diag}
\newcommand*{\Scale}[2][4]{\scalebox{#1}{$#2$}}

\def\idty{{\mathchoice {\mathrm{1\mskip-4mu l}} {\mathrm{1\mskip-4mu l}} %
{\mathrm{1\mskip-4.5mu l}} {\mathrm{1\mskip-5mu l}}}}

\title[Localization in the XY Chain]{Localization Properties of the Disordered XY Spin Chain}
\subtitle{A review of mathematical results with an eye toward Many-Body Localization}

\author[H. Abdul-Rahman]{Houssam Abdul-Rahman\inst{1}}
\author[B. Nachtergaele]{Bruno Nachtergaele\inst{2}}
\author[R. Sims]{Robert Sims\inst{1}}
\author[G. Stolz]{G\"unter Stolz\inst{3}\footnote{Corresponding author\quad E-mail:~\textsf{stolz@math.uab.edu}}}
\address[1]{Department of Mathematics\\
University of Arizona\\
Tucson, AZ 85721, USA}
\address[2]{Department of Mathematics\\ University of California, Davis\\
Davis, CA 95616. USA}
\address[3]{Department of Mathematics\\
University of Alabama at Birmingham\\
Birmingham, AL 35294 USA}
\shortauthors{H. Abdul-Rahman et al.}
\begin{abstract}
We review several aspects of Many-Body Localization-like properties exhibited by the
disordered XY chains: localization properties of the
energy eigenstates and thermal states, propagation bounds of Lieb-Robinson type, decay of correlation functions, absence of
particle transport, bounds on the bipartite entanglement, and bounded entanglement growth under the dynamics. We also prove
new results on the absence of energy transport and Fock space localization. All these properties are made accessible to mathematical
analysis due to the exact mapping of the XY chain to a system of quasi-free fermions given by the Jordan-Wigner transformation.
Motivated by these results we discuss conjectured properties of more general disordered quantum spin and other systems as possible
directions for future mathematical research.
\end{abstract}
\shortabstract
\begin{document}
\maketitle

\section{Introduction}\label{sec:intro}

\subsection{The many-body localization context}

Quantum spin systems provide a promising class of models in the quest for a better understanding of many-body localization (MBL). In particular, spins (as opposed to, say, electrons) are zero-dimensional and have essentially trivial one-body dynamics, so that studying systems of interacting spins allows to fully focus on many-body effects and thus provides a clear view on the various phenomena which are physically associated with MBL. While studying MBL in systems of interacting fermions remains a long-term goal, in the meantime quantum spin systems provide an ideal laboratory to help identifying the relevant effects characterizing MBL and how they manifest themselves.

The most recent decade has seen extensive research on MBL in the physics literature. A key insight at the beginning of much of this work was the understanding that MBL should be understood as localization in Fock space, which provides a means to describe that the eigenstates of an interacting many-body system in the MBL phase should arise as perturbations of the eigenstates of the corresponding non-interacting system. This point of view was stressed in \cite{Baskoetal} and \cite{Gornyietal}, as well as earlier work in \cite{Altshuleretal}. We will return to this below.

Some of the works which followed and laid further groundwork for describing the many-body localization transition are
\cite{OganesyanHuse, Znidaricetal, PalHuse, HuseOganesyan, VoskAltman2013}. These and other works also shifted focus to using quantum spin systems as models for understanding MBL. We will point to additional physics references in more concrete contexts below, but do not attempt to provide a comprehensive list. Instead, we refer to other survey articles, to be published in the same volume as this article, for a more complete discussion of the current state of knowledge on MBL in theoretical and experimental physics.

In this work our main goal is to describe a broad range of MBL-like properties of the disordered XY chain, where fully rigorous proofs can be given. This is due to the fact, going back to \cite{LSM}, that the XY chain, via the Jordan-Wigner transform, can be reduced to a free Fermion system. Thus the XY chain becomes exactly solvable in the translation invariant case. For variable coefficients, the case relevant to us here, the Jordan-Wigner transform reduces the XY chain to the study of an effective one-particle Hamiltonian. Thus properties of the one-particle Hamiltonian translate to properties of the many-body spin system, where some attention has to be given to locality issues (as the Jordan-Wigner transform is non-local).

If, for example, a translation invariant XY chain is subjected to a random transversal magnetic field, then the effective Hamiltonian becomes the Anderson model. As a consequence, known localization properties of the Anderson model can be used to rigorously prove localization properties of the XY chain in a random field. Here we will discuss a broad range of results of this type. In particular, we will describe recent localization results in the disordered XY chain and closely related disordered free Fermion systems from \cite{HSS, SimsWarzel, PasturSlavin, AR-S, ANSS}. In the last two subsections we will also include several previously unpublished results, where we will provide detailed proofs.

Many theoretical physicists believe that MBL is a qualitatively different phenomenon from one-body localization in a many-particle system.  In particular, due to the possibility of mapping to a free Fermion system, localization properties of the disordered XY chain are often viewed as a form of Anderson localization as opposed to MBL. This may be true but we believe that the extent of this distinction can only be properly understood by studying the detailed properties of both one-body and many-body localization in specific models. One should keep in mind that many-body effects in an interacting system can often be qualitatively approximated by an effective one-body model, such as is often done in mean-filed theory or Fermi liquid theory. Another example is the discovery that the Haldane phase represented by the ground state of the AKLT chain is in fact the same phase (in the sense of \cite{chen:2011}) that contains models with a unique (up to edge states) product ground state \cite{schuch:2011,bachmann:2014}. It is our view that, whatever the precise relationship turns out to be, studying the dynamical and response properties relevant for MBL in disordered models such as the random XY chain will be helpful.

\subsection{Results surveyed in this work}

Sections~\ref{sec:XY} and \ref{sec:onepart} briefly recall the reduction of the XY chain, via the Jordan-Wigner transform, to a free Fermion system and discuss eigencorrelator localization for the associated effective Hamiltonian. Throughout our work we will use this well understood form of one-particle localization to derive MBL properties of the XY chain.

Among the MBL results covered is absence of many-body transport in the form of vanishing group velocity (i.e.\ a so-called zero-velocity Lieb-Robinson bound), see Section~\ref{sec:LR}. Following earlier discussion in \cite{BO}, this was rigorously shown in \cite{HSS}.

In Section~\ref{sec: correlations} we present results of \cite{SimsWarzel}, showing that {\it all} eigenstates of the disordered XY chain, as well as thermal states at any inverse temperature, have exponentially decaying correlations. In this context we stress that, in order to be interpreted as an MBL property, it is important to be able to go beyond ground state correlations, as MBL should reflect the existence of a mobility gap without the need of a ground state gap. In fact, for the disordered XY chain the mobility gap extends through the entire spectrum, i.e.\ there is no mobility edge. 

We mention that there are general results which show that zero-velocity Lieb-Robinson bounds imply exponential decay of eigenstate correlations \cite{HSS, Friesdorfetal}. However, these results either give weaker bounds or use additional assumptions which do not hold in the disordered XY chain (e.g.\ the relevant constants can not be chosen uniform in the random parameters, see also the discussion is Section~\ref{sec:remarks} below). They also do not apply to thermal states. For the XY chain, the direct analysis carried out in \cite{SimsWarzel} gives the best bounds.

In Section~\ref{sec:arealaws} we discuss area laws for the bipartite entanglement of eigenstates. The validity of such bounds at macroscopic energies (going beyond the gapped ground state phase where area laws were first identified in \cite{Hastings}) is considered another characteristic of MBL, see e.g.\ \cite{BauerNayak} and \cite{BrandHorNP}. For the disordered XY chain a uniform area law for all eigenstates was shown in \cite{AR-S}, see (\ref{statent}) in Section~\ref{sec:arealaws}. As we will describe, this is based on the fact, shown in \cite{PasturSlavin}, that eigenstates of disordered free Fermion systems in arbitrary dimension $d$ satisfy an area law bound. In dimension $d=1$ this translates to an entanglement bound for the XY chain. In Section~\ref{sec:arealaws} we will also discuss the relation of the entanglement bounds found for the XY chain with a recent general result by Brandao and Horodecki \cite{BrandHorNP}.

Similar methods can also be used to prove an area law for the dynamical entanglement, i.e.\ that the entanglement entropy of an initial product state stays uniformly bounded in time. Bounds on the dynamical entanglement in disordered spin chains were previously discussed in \cite{Bardarsonetal, Serbynetal2013a, VoskAltman2014, ho:2015}. A rigorous result for the XY chain, proven in \cite{ANSS}, is presented as (\ref{dynent}) in Section~\ref{sec:arealaws} below.

A related open problem is to better understand entanglement bounds for thermal states and, in particular, getting such bounds on efficient entanglement measures of thermal states such as the logarithmic negativity. However, an area law uniform in all eigenstates  easily implies an area law for the entanglement of formation of thermal states, see (\ref{entform}).

Results on the absence of particle transport and energy transport are discussed in Section~\ref{sec:parttrans}. Here we will focus mostly on the isotropic XY chain, which preserves the number of up-spins or `particles' (in the interpretation as a free Fermion system), thus leading to stronger results. The results on energy transport, Theorems~\ref{thm:energytrans} and \ref{thm:energytrans2} have not been published before. Thus we state them more formally and include a detailed proof in Appendix~\ref{appA}.

Finally, in Section~\ref{sec:fock}, we return to the concept of Fock space localization. As discussed in several of the references given above, this concept is emerging as the central characteristic of MBL, leading to many of the manifestations noted above as secondary consequences. In this context, our main result for the disordered XY chain is part (a) of Theorem~\ref{thm:fock}, which shows that the eigenstates of the interacting spin chain are close, in suitable sense, to the eigenstates of the non-interacting system. As a distance measure we choose distance of Fock space configurations. Part (b) of Theorem~\ref{thm:fock} illustrates this, with a simpler proof, in terms of local spin occupation numbers. These are new results for which we provide proofs in Appendix~\ref{proof8}.

In the concluding summary section we mention some conjectures and possible directions for future mathematical research.

\subsection{The limits of localization effects} \label{sec:remarks}

We chose here to present our results in terms of bounds on averages of the relevant quantities. Such bounds are not only what is considered in most numerical studies, 
but they also have proven to be best suited for mathematical investigations and give the most concise and to-the-point descriptions of the results. As all the random quantities considered are non-negative, averages describe their typical behavior to leading order.

It is certainly desirable, in particular when attempting to understand the associated random variables beyond leading order and to study their fluctuations, to also have bounds on the probability that a system satisfies one of the localization properties discussed here. An easy way to get such bounds from expectation bounds is via Chebychev's inequality. Say that a non-negative random quantity $A_{xy}$ has exponentially decaying expectation $\E(A_{xy}) \le Ce^{-\xi|x-y|}$. Then $\mathbb{P}(A_{xy} \le e^{-\xi|x-y|/2}) \ge 1 - Ce^{-\xi|x-y|/2}$.

In a system of finite size, e.g.\ the length $n$ of a finite spin chain in our setting, such quantitative localization bounds obviously do not hold with probability one, as there will be realizations of the random configuration where the system is close to translation invariant and localization can't be expected. What is more important is that probability bounds can not be independent of $n$ as the system size increases and, in particular, will not yield useful quantitative bounds in the thermodynamic limit $n\to\infty$. 

An example which demonstrates this are the bounds on eigenfunction localization in the Anderson model which we discuss in Lemma~\ref{lem:etaloc} below as well as the resulting Fock space localization result for the XY chain in Theorem~\ref{thm:fock}. Here the probabilities as well as the actual localization bounds pick up some power-law dependence on $n$. As we explain there, this is not merely an artifact of the proof and due to the fact that large random systems will, with high probability, have large subsystems which are essentially translation invariant.

As another example we refer to the area laws for the bipartite entanglement of eigenstates of the XY chain discussed in Section~\ref{sec:arealaws}. When considering such bounds in the thermodynamic limit $n\to\infty$ for a subsystem of size $\ell$, then averages of the bipartite entanglement of eigenstates remain bounded uniformly in $\ell$, but by a result in \cite{EPS} the entanglement is not self-averaging in the sense that with probability one it is unbounded in $\ell$. Thus there is no area law in probability. The reason for this is again that one-dimensional disordered systems have large subsystems which are almost translation invariant. If the subsystem boundary falls into one of these regions, it will lead to large entanglement.

It is for these reasons that using expectations has led to the more satisfying mathematical results on localization, which, as we feel, also are best suited to convey the physical meaning of localization effects. Note that, as a technical consequence, this means that the proofs of the results described below are of the form ``expectation bounds for the effective one-particle Hamiltonian imply expectation bounds for quantities related to the spin system''. This needs somewhat different strategies of proof then results where a deterministic input leads to a deterministic output.

\section{XY chain and effective Hamiltonian} \label{sec:XY}

A general XY spin chain in transversal field is described by the Hamiltonian
\begin{equation} \label{XYchain}
H = - \sum_{j=1}^{n-1} \mu_j ((1+\gamma_j) \sigma_j^X \sigma_{j+1}^X + (1-\gamma_j)\sigma_j^Y \sigma_{j+1}^Y) - \sum_{j=1}^n \nu_j \sigma_j^Z,
\end{equation}
where $\sigma_j^X$, $\sigma_j^Y$ and $\sigma_j^Z$ are the standard Pauli matrices acting on the $j$-th spin. The parameter sequences $\mu_j$, $\gamma_j$ and $\nu_j$, describing the interactions strength, anisotropy and field strength, are viewed as random parameters reflecting the disorder in the system. We will generally assume that all three sequences are independent and identically distributed, with bounded distribution, although this could be weakened in several ways (as long as the required localization properties of the effective Hamiltonians, see below, can be shown).

We will state all results for the finite XY chain, i.e.\ restricted to the interval $\Lambda = [1,n]$, although with constants bounded uniformly in $n$. This allows to draw conclusions for the infinite chain, although we will not disucss this here.

The Jordan-Wigner transform 
\[c_1 = a_1,\ c_j = \sigma_1^Z \ldots \sigma_{j-1}^Z a_j,\ j=2,\ldots,n,\] with the spin lowering operators $a_j = (\sigma_j^X-i\sigma_j^Y)/2$, satisfies the canonical anti-commutation relations (CAR)  $\{c_j, c_k^*\} = \delta_{jk}$, $\{c_j, c_k\} = \{c_j^*, c_k^*\}=0$, and maps the XY chain onto a system of quasi-free Fermions governed by
\begin{equation} \label{JWrep}
H = \sum_{j,k} \left(c_j^* A_{jk} c_k - c_j A_{jk} c_k^* + c_j^* B_{jk} c_k^* - c_j B_{jk} c_k\right)
\end{equation}
Here $A$ and $B$ are the tridiagonal $n\times n$-matrices
\begin{eqnarray} \label{A}
A &=& \begin{pmatrix} -\nu_1 & \mu_1 & & \\ \mu_1 & \ddots & \ddots & \\ & \ddots & \ddots & \mu_{n-1} \\ & & \mu_{n-1} & -\nu_n \end{pmatrix}, \\ B &=& \begin{pmatrix} 0 & \mu_1\gamma_1 & & \\ -\mu_1\gamma_1 & \ddots & \ddots & \\ & \ddots & \ddots & \mu_{n-1}\gamma_{n-1} \\ & & -\mu_{n-1} \gamma_{n-1} & 0 \end{pmatrix}. \label{B}
\end{eqnarray}
If $\gamma_j =0$, $j \in \Lambda$, i.e.\ for the isotropic XY chain, we have $B=0$ and the matrix $A$ has the role of an effective Hamiltonian. In the general anisotropic case the effective Hamiltonian is the $2n\times 2n$-matrix $\begin{pmatrix} A & B \\ -B & -A \end{pmatrix}$. However, to correctly reflect locality properties, we re-index this matrix and write the spin Hamiltonian as the quadratic form
\begin{equation} \label{quadform}
H={\mathcal C}^* M {\mathcal C}
\end{equation} in terms of the column vector ${\mathcal C} = (c_1,c_1^*, \ldots, c_n, c_n^*)^t$ and effective Hamiltonian given as the $2\times 2$-block Jacobi matrix
\begin{equation} \label{effHamanisotropic}
\Scale[0.95]{\displaystyle M = \left( \begin{array}{cccc} -\nu_1 \sigma^Z & \mu_1 S(\gamma_1) & & \\ \mu_1 S(\gamma_1)^t & \ddots & \ddots & \\ & \ddots & \ddots & \mu_{n-1} S(\gamma_{n-1}) \\ & & \mu_{n-1} S(\gamma_{n-1})^t & -\nu_n \sigma^z \end{array} \right).}
 \end{equation}
Here $\sigma^Z = \begin{pmatrix} 1 & 0 \\ 0 & -1 \end{pmatrix}$ is the third Pauli matrix and $S(\gamma) = \begin{pmatrix} 1 & \gamma \\ -\gamma & -1 \end{pmatrix}$.

We note that, instead of the fermionic modes $c_j$ and $c_j^*$, commonly used are also the Majorana fermions $a_j^+ = c_j^*+c_j$, $a_j^- = i(c_j^*-c_j)$, which lead to a slightly different (but unitarily equivalent) form of the effective Hamiltonian.

The quadratic Hamiltonian (\ref{quadform}) can be diagonalized via a Boguliubov transformation implemented by a $2n\times 2n$ Bogoliubov matrix $W$ (meaning that $W$ is orthogonal with $WJW^t = J$ for $J=(\sigma^X)^{\oplus n}$). $W$ diagonalizes $M$ as
\begin{equation} \label{diagM}
WMW^t = \bigoplus_{j=1}^n \begin{pmatrix} \lambda_j & 0 \\ 0 & -\lambda_j \end{pmatrix},
\end{equation}
with $0\le \lambda_1 \le \ldots \le \lambda_n$ (which can be identified as the singular values of $A+B$, e.g.\cite{HSS}).
This gives a new set of fermionic modes $b_j$, $b_j^*$, represented in vector form as ${\mathcal B} = W {\mathcal C}$, which expresses $H$ as a free fermion system,
\begin{equation}
H = 2 \sum_{j=1}^n \lambda_j b_j^* b_j - E_0 \idty,
\end{equation}
with $E_0 = \sum_j \lambda_j$. The Fock basis
\begin{equation} \label{eigenstates}  \psi_{\alpha} = \prod_{j=1}^n (b_j^*)^{\alpha_j} \psi_0, \quad \alpha \in \{0,1\}^n,
\end{equation}
with respect to the vacuum vector $\psi_0$ of this system gives a full set of eigenvectors of $H$ with corresponding eigenvalues $\sum_{j:\alpha_j =1} \lambda_j - E_0$.

\section{One-Particle Localization} \label{sec:onepart}

The phenomena of single-particle localization and many-body localization should be clearly distinguished. The latter describes properties of group waves as opposed to waves associated with individual particles in a many-body system. For the XY chain, however, MBL properties of $H$ are consequences of single-particle localization properties of the effective Hamiltonian $M$ (but note that $M$ does not act on a spin, but on a particle with an infinite degree of freedom as $n\to\infty$, e.g.\ a one-dimensional `electron').

While this is the main mechanism behind the results presented here, we stress that it makes the XY chain a rather unrealistic example of a many-body quantum system. One other example which shares the property of being fully reducible to an effective one-particle Hamiltonian are interacting harmonic quantum oscillators, see \cite{NSS1,NSS2}. For other classes of disordered quantum many-body systems the relation between one-particle and many-body localization, and the variety of possible MBL regimes and phenomena, will likely be much more involved.

We will state results under the assumption of exponential {\it eigencorrelator localization} for $M$, i.e.\ the existence of a finite constant $C$ and a positive constant $\eta$, not depending on $n$, such that
\begin{equation} \label{eq:eigcorloc}
\E \Big( \sup_{|g|\le 1} \|g(M)_{jk}\| \Big) \le C e^{-\eta|j-k|}
\end{equation}
for all $1\le j,k \le n$. Here $\E(\cdot)$ denotes disorder averaging and $g(M)$ is defined via the spectral theorem for symmetric matrices, i.e.\ $g(M) = \sum_j g(\lambda_j) |\phi_j\rangle \langle \phi_j|$, where $\phi_j$ and $\lambda_j$ are a full system of eigenvectors and corresponding eigenvalues for $M$. For later convenience, we think of the $2n\times 2n$-matrix $g(M)$ and an $n\times n$-matrix with $2\times 2$-matrix-valued entries $g(M)_{jk}$ and of $\|\cdot\|$ as a $2\times 2$-matrix norm.

In the $n\to\infty$ limit, eigencorrelator localization (\ref{eq:eigcorloc}) implies that $M$ has pure point spectrum with exponentially decaying eigenfunctions (with $\eta$ giving the inverse localization length), but is a mathematically stronger property. In particular, with the choice $g(M) = e^{-itM}$, it includes uniform exponential decay of time evolution amplitudes \begin{equation} \label{eq:dynloc}
\E \Big( \sup_{t\in \R} \|(e^{-itM})_{jk}\| \Big) \le C e^{-\eta|j-k|}.
\end{equation}
Also covered by (\ref{eq:eigcorloc}) are the Fermi projectors $\chi_{(-\infty,E]}(M)$, i.e.\ eigenprojectors onto energies below $E$. Eigencorrelator localization in the above strong form allows to derive all the MBL properties to be discussed below.

For the isotropic case, $M$ can be replaced by $A$ in (\ref{eq:eigcorloc}) and is known to hold for large classes of random parameters $\nu_j$ and $\mu_j$. In particular, if $\mu_j=1$ and the $\nu_j$ are i.i.d.\ with sufficiently smooth distribution (for example if they have bounded compactly supported density), then $A$ is the one-dimensional Anderson model and (\ref{eq:eigcorloc}) is known since the work of Kunz and Souillard \cite{KS} and, for larger classes and under weaker assumptions, can also be proven using the fractional moments method of Aizenman and Molchanov \cite{AM}. See \cite{Stolz,AW} for general introductions to the mathematical theory of Anderson localization.

Proving (\ref{eq:eigcorloc}) in the anisotropic case is mathematically more challenging due to the lack of monotonicity of (\ref{effHamanisotropic}) in the random parameters. A general result of \cite{Elgartetal} covers the case of a magnetic field at large disorder, i.e.\ $\nu_j$ is multiplied with a sufficiently parameter $\lambda$.  For additional results see \cite{ChapmanStolz}.

For this work we use the assumption of {\it exponential} decay of eigencorrelators mostly for ease of presentation and for its physical relevance. Many of the results discussed below, in suitably modified form, also hold under weaker assumptions on eigencorrelator decay, such as sufficiently fast power law decay.

Another assumption needed for some of the results below is that
\begin{equation} \label{eq:nondeg}
\mbox{\hspace{.7cm} \parbox[c]{6.5cm}{all eigenvalues of $H_n$ are non-degenerate
for almost every realization of the disorder.}}
\end{equation}
This is true, for example, if the random variables $\nu_j$ have continuous distribution, see Appendix A in \cite{AR-S}.

\section{Zero-velocity Lieb-Robinson bounds} \label{sec:LR}

A local observable acting on the $j$-th spin via the $2\times 2$-matrix $A$ will be denoted by $A_j$. For $j\not= k$ local observables commute: $[A_j, B_k] = 0$. Lieb-Robinson bounds, initially introduced in \cite{LR}, provide upper bounds on the propagation speed of group waves through the spin system by bounding the norm of the commutators under the time-evolution $\tau_t(A_j) = e^{itH} A_j e^{-itH}$ of one of the observables. In general, a Lieb-Robinson bound of the form
\begin{equation}  \|[ \tau_t(A_j), B_k ]\| \le C \|A\| \|B\| e^{-\eta (|j-k|-vt)},
\end{equation}
with constants $C$ and $\eta>0$ which do not depend on $n$, $j$, $k$ and $t$ and hold for all local observables $A$ and $B$, show that $v$ is an upper bound on the group velocity in the system, see \cite{NS2009} for a survey of recent results and applications.

It was shown in \cite{HSS} that the introduction of disorder into the $XY$-chain leads to Lieb-Robinson velocity zero in the disorder average: Under the one-particle localization condition (\ref{eq:eigcorloc}) (in fact (\ref{eq:dynloc}) suffices here) it holds that
\begin{equation} \label{LRbound}
\E \left( \sup_{t\in \R} \|[\tau_t(A_j), B_k] \| \right) \le C'\|A\| \|B\| e^{-\eta|j-k|},
\end{equation}
uniformly in $n$ and $1\le j,k \le n$.

For the case of the isotropic $XY$-chain in random transversal field a slightly weaker form of this bound, with right hand side growing linearly in $t$ and quadratically in $n$, was discussed earlier in \cite{BO}. The key fact behind the proof of (\ref{LRbound}) is the relation
\begin{equation} \label{dynrelation}
\tau_t({\mathcal C}) = e^{-2itM} {\mathcal C},
\end{equation}
where the time evolution of the vector $\mathcal C$ is understood componentwise. This relates the one-particle dynamics of $M$, given by the $2n \times 2n$-matrix $e^{-2itM}$, to the many body dynamics of $\mathcal C$ and thus, through Jordan-Wigner, of local observables in the spin chain. Here some care is needed to account for the non-locality of the Jordan-Wigner transform: Assume $k\ge j$, then (\ref{eq:dynloc}), (\ref{dynrelation}) and a geometric summation give
\begin{equation}
\E \left( \sup_t \|[\tau_t(c_j), B_k ]\| \right) \le \frac{4C\|B\|}{1-e^{-\eta}} e^{-\eta(k-j)}.
\end{equation}
Writing the local spin lowering operators as $a_j = \sigma_1^Z \ldots \sigma_{j-1}^Z c_j$, an iterative argument based on the Leibnitz rule followed by another geometric sum gives
\begin{equation}
\E \left( \sup_t \|[\tau_t(a_j), B_k ]\| \right) \le \frac{16C\|B\|}{(1-e^{-\eta})^2} e^{-\eta(k-j)}.
\end{equation}
Using that $a_j$, $a_j^*$, $a_j^* a_j$ and $a_j a_j^*$ generate the local operators at site $j$, this leads to the claim (\ref{LRbound}) with $C' = 96C/(1-e^{-\eta})^2$.
More details on this proof are in \cite{HSS}; see also the review \cite{sims:2015}.

More recent results of this type for disordered systems
can be found in \cite{geb:2016}, which establishes an MBL transition for the XY chain in decaying (and thus non-ergodic) random field in the $z$-direction. Related bounds for quasi-periodic models are contained in e.g. \cite{dam:2014}, \cite{kach:2015}, and references therein.

\section{Rapid decay of correlations} \label{sec: correlations}

One of the benefits of Lieb-Robinson bounds (with finite velocity $v$) is that they can be used to prove exponential clustering of the ground state for {\it gapped} spin systems, i.e.\ exponential decay of the ground state correlations
\begin{equation}
|\langle \psi_0, A_j B_k \psi_0 \rangle - \langle \psi_0, A_j \psi_0 \rangle \langle \psi_0, B_k \psi_0 \rangle|
\end{equation}
in $|j-k|$ for local observables $A_j$ and $B_k$, see \cite{Hastings2004, NS2006}.

In \cite{HSS} it was shown that in the presence of a zero-velocity Lieb-Robinson bound the assumption of a uniform ground state gap can be relaxed, leading to a correction of the exponential clustering bound which is logarithmic in the inverse gap size. For the case of the isotropic $XY$-chain in random transverse field this leads to a bound of the form $C\|A\| \|B\| n e^{-\eta |j-k|}$ for the disorder average of ground state correlations (Theorem~4.2 in \cite{HSS}). The dependence on the length $n$ of the chain is due to the use of a Wegner estimate for the Anderson model, giving a volume dependent ground state gap. For earlier work on this question, employing bounds on the Anderson model found by multiscale analysis, see \cite{KleinPerez}.

We stress that rapid decay of ground state correlations, by itself, is not sufficient to indicate MBL, as it holds for many translation invariant systems, even without requiring a spectral gap \cite{N,FSWCP}. Rather, MBL should reflect a {\it mobility gap}, i.e.\ a range of energies near the bottom of the spectrum which is entirely many-body localized. For correlations in the disordered XY chain this was recently settled in full generality by Sims and Warzel \cite{SimsWarzel}. They consider time dependent correlations
\begin{equation} \label{dyn_cor}
 \langle \tau_t(A_j) B_k \rangle - \langle A_j \rangle \langle B_k \rangle
 \end{equation}
for the anisotropic disordered $XY$-chain (\ref{XYchain}), where $\langle \cdot \rangle = \tr (\rho \,\cdot)$ and $\rho = |\psi_{\alpha} \rangle \langle \psi_{\alpha}|$ for an arbitrary eigenstate $\psi_{\alpha}$ of $H$ or a thermal state $\rho = e^{-\beta H}/ \tr e^{-\beta H}$. Assuming eigencorrelator localization (\ref{eq:eigcorloc}) it is shown that
\begin{equation} \label{ave_cor_dec}
\E \left( \sup_{t\in \R} | \langle \tau_t(A_j) B_k \rangle - \langle A_j \rangle \langle B_k \rangle | \right) \le C e^{-\eta|j-k|},
\end{equation}
where the constants $C<\infty$ and $\eta>0$ can be chosen uniformly in the eigenstate label $\alpha$ and the inverse temperature $\beta>0$, respectively. In fact, having exponential clustering for all eigenstates means that the disordered XY chain in fully many-body localized at all energies, i.e.\ in the infinite temperature limit.

Let us now discuss a few of the central ideas that are used in proving (\ref{ave_cor_dec}).
In order to do so, we first review some basics about quasi-free states on the CAR algebra
and refer the interested reader to e.g. \cite{Brat_Rob2} for more details.
Recall that for any Hilbert space $\mathcal{H}$, one can associate to each
$f \in \mathcal{H}$ annihilation and creation operators, which we label by $c(f)$ and $c^*(f)$ respectively.
These operators act on the fermionic (i.e. anti-symmetric) Fock space $\mathcal{F}( \mathcal{H})$ corresponding to $\mathcal{H}$
and satisfy canonical anti-commutation relations (CAR), i.e.
\begin{eqnarray}
\{ c(f), c(g) \} &=& \{ c^*(f), c^*(g) \} = 0 \quad \mbox{and} \nonumber \\ \quad \{ c(f), c^*(g) \} &=& \langle f, g \rangle \idty \quad \mbox{for all } f,g \in \mathcal{H} \,
\end{eqnarray}
with $\{ A, B \} = AB+BA$. The $C^*$-algebra $\mathcal{A}( \mathcal{H})$ generated by the identity $\idty$ and the operators $c(f)$ and $c^*(g)$ for all $f,g \in \mathcal{H}$ is called the CAR algebra associated to $\mathcal{H}$.

A state $\omega$ on $\mathcal{A}( \mathcal{H})$ is said to be quasi-free if all its correlation
functions can be computed using Wick's rule. An important sub-class of
quasi-free states $\omega_{\varrho}$ on $\mathcal{A}( \mathcal{H})$ are uniquely determined
by a one-particle density operator $0 \leq \varrho \leq \idty$ acting on $\mathcal{H}$.
In this case, multipoint correlation functions corresponding to these states
have a determinantal structure, e.g.
\begin{equation} \label{multi_cor}
\Scale[0.85]{\displaystyle\omega_{\varrho} \left( c^*( g_m) \cdots c^*( g_1) c(f_1) \cdots c( f_m) \right) = {\rm det} \left( \langle f_j, \varrho g_k \rangle \right)_{1 \leq j,k \leq m}}
\end{equation}
for any $m \geq 1$ and $f_1, \cdots, f_m, g_1, \cdots, g_m \in \mathcal{H}$. Both classes
of states considered in (\ref{ave_cor_dec}), i.e. eigenstates and thermal states of the XY-model,
are quasi-free states on $\mathcal{A}( \ell^2([1,n]))$ with this particular form. We remark that a
full characterization of quasi-free states on $\mathcal{A}(\mathcal{H})$ can be found in \cite{Bachnfriends},
but this goes beyond the scope of our applications.

A key result in \cite{SimsWarzel} estimates structured determinants; in particular, the result applies to
multi-point correlation functions associated to these quasi-free states, see (\ref{multi_cor}), in the case that $\mathcal{H}= \ell^2( \mathbb{Z})$. (The result also applies in the
case of $\mathcal{H} = \ell^2([1,n])$ for any $n \geq 1$.) To state a version of this
result, first introduce for $m \geq 1$, an ordered configuration as ${\rm x} = (x_1, \cdots, x_m) \in \mathbb{Z}^m$ with $x_1 < \cdots < x_m$. For any two such configurations ${\rm x}$ and
${\rm y}$, denote a configuration distance by
\begin{equation} \label{config_dist}
D( {\rm x}, {\rm y}) = \max_{1 \leq j \leq m} |x_j - y_j|.
\end{equation}
The following is proven in \cite{SimsWarzel}.
\begin{thm} \label{thm:SW} Let $K:[0, \infty) \to [0, \infty)$ be monotone increasing and suppose there is some $\mu_0 \in (0, \infty)$
for which
\begin{equation}
I( \mu_0) := \sum_{\ell =0}^{\infty} (1+ \ell)e^{- \mu_0 K( \ell)} < \infty
\end{equation}
For each $t \in \mathbb{R}$, let $\rho(t)$ be an operator on $\ell^2( \mathbb{Z})$ with $\| \rho(t) \| \leq 1$.
If there is some $C < \infty$ and $\mu > \mu_0$ for which given any $x,y \in \mathbb{Z}$, one has the estimate
\begin{equation} \label{pair_dec}
\sup_{t \in \mathbb{R}} | \langle \delta_x, \rho(t) \delta_y \rangle| \leq C e^{- \mu K(|x-y|)} \, ,
\end{equation}
then for any $m \geq 1$ and any pair of ordered configurations ${\rm x}$ and ${\rm y}$, one has that
\begin{equation}\label{det_dec}
\Scale[0.9]{\displaystyle\sup_{t \in \mathbb{R}} \left| {\rm det} ( \langle \delta_{x_j}, \rho(t) \delta_{y_k} \rangle )_{1 \leq j,k \leq m} \right| \leq  C'
{\rm exp}\left( - \frac{\mu-\mu_0}{2} K\left( \frac{D({\rm x}, {\rm y})}{2} \right) \right)}
\end{equation}
Here one may take $C' = 8 \max\{ CI(\mu_0), \sqrt{CI( \mu_0)} \}$.
\end{thm}

In words, this result shows that decay in the entries of a matrix, i.e. (\ref{pair_dec}), does lead to a form of decay in the
corresponding matrix determinant, see (\ref{det_dec}). The decay in the determinant is expressed in terms of the
configuration distance $D({\rm x}, {\rm y})$, see (\ref{config_dist}).

With these preliminaries, we can now give some rough ideas for how (\ref{ave_cor_dec}) is proven.
Consider a quasi-free state $\omega_{\varrho}$ on $\mathcal{A}( \ell^2([1,n]))$ of the
form described in (\ref{multi_cor}) above and let $\tau_t$ denote the dynamics corresponding
to the isotropic XY-model. For any pair of ordered configurations ${\rm x} =(x_1, \cdots, x_m)$
and ${\rm y} =(y_1, \cdots, y_m)$, with components in $[1,n]$, one has
\begin{eqnarray} \label{dyn_cor_qfs}
\omega_{\varrho} \left( \tau_t(c^*( \delta_{y_m})) \cdots \tau_t( c^*( \delta_{y_1})) c(\delta_{x_1}) \cdots c( \delta_{x_m}) \right) \nonumber \\ =
{\rm det} \left( \langle \delta_{x_j}, \varrho e^{-2itA} \delta_{y_k} \rangle \right)_{1 \leq j,k \leq m}
\end{eqnarray}
where $A$ is the effective one-particle Hamiltonian and $t \in \mathbb{R}$ is arbitrary. Given the assumptions of exponential eigencorrelator localization and non-degeneracy, i.e. (\ref{eq:eigcorloc}) and (\ref{eq:nondeg}), one concludes that the matrix entries on the right-hand side of (\ref{dyn_cor_qfs}) decay
and as a result, these multi-point correlations decay as well, by an application of Theorem~\ref{thm:SW}.
A technical point here is that the assumption of exponential eigencorrelator
localization only guarantees that the disorder-averaged matrix entries, found on the right-hand side of (\ref{dyn_cor_qfs}), decay.
However, a further result in \cite{SimsWarzel}, see Theorem 1.2 therein, proves a generalization of Theorem~\ref{thm:SW} which covers this random situation. In words,
the proof shows that if the averaged matrix entries decay similarly to (\ref{pair_dec}), then the averaged determinant decays similarly
to (\ref{det_dec}) In particular, both results obtain estimates which are uniform in $m, n,$ and $t \in \mathbb{R}$.

Now, using the Jordan-Wigner transform, the dynamic correlations considered in (\ref{dyn_cor})
can be re-written in terms of time-dependent multi-point correlations, similar to those on the left-hand-side of (\ref{dyn_cor_qfs}),
in specific quasi-free states with the form discussed above. In full generality, the evaluation of these multi-point correlations,
i.e. the application of Wick's rule, results in a pfaffian; not a determinant. Since pfaffians and determinants share similar
mathematical properties, it is not surprising that analogs of the theorems mentioned above hold
in this case as well. Such technical matters are discussed in detail and proven in \cite{SimsWarzel}.
It is with these ideas that (\ref{ave_cor_dec}) is proven.

\section{Area laws for the bipartite entanglement} \label{sec:arealaws}

Another property which reflects many-body localization are area laws for the bipartite entanglement of eigenstates and for the time evolution of initially unentangled states, e.g.\ \cite{Bardarsonetal, BauerNayak, BrandHorNP, BrandHor15, ho:2015, Serbynetal2013a}. In fact, such bounds are closely related to the rapid decay of correlations discussed in the previous section. Hastings showed in his seminal work \cite{Hastings} that the ground state of a uniformly gapped one-dimensional spin system satisfies an area law, i.e.\ its bipartite entanglement with respect to a subchain is uniformly bounded, independent of the length of the chain and the subchain. As mentioned above such systems also satisfy exponential clustering of the ground state. In fact, Brandao and Horodecki \cite{BrandHorNP, BrandHor15} proved a general result saying that exponential decay of correlations of one-dimensional quantum states implies an area law. This does not only provide a new proof of Hastings' result, but also applies in cases where exponential decay of correlations is known for excited states, without knowledge on spectral gaps, and thus is applicable to the situation one wants to see in MBL. In particular, it is likely that the results of \cite{HSS} and \cite{SimsWarzel} on correlation decay in the disordered XY chain combine with the result of \cite{BrandHor15} to provide area laws for this example. Apart from the fact that  in \cite{BrandHor15} the authors work with periodic boundary conditions, a potential obstacle arises from the fact that the correlation bounds required in \cite{BrandHor15} are stronger than those provided in \cite{SimsWarzel}. The issue is, once again, the prefactor. The theorem in \cite{BrandHor15} assumes not only a uniform correlation length but also that the prefactor is just the product of the norms of two arbitrary observables and neither depends on the size of the system nor on the size of the supports of the two observables.

This is one of the reasons why it is desirable to have a more direct argument for an area law of the eigenstate entanglement in the example of the disordered XY chain. Towards this, decompose the chain $\Lambda$ into left and right ends $A=[1,\ell]$ and $B=[\ell+1,n]$. For a pure state $\rho = |\psi \rangle \langle \psi|$, let $\rho_A$ and $\rho_B$ be the corresponding reduced states. The bipartite entanglement of $\rho$ with respect to this decomposition is the von Neumann entropy of the reduced state,
\begin{equation} \label{traceformula}
{\mathcal E}(\rho) = {\mathcal S}(\rho_A) = - \tr \rho_A \log \rho_A.
\end{equation}
Generic states will satisfy the volume law ${\mathcal E}(\rho) \sim \min\{\ell,n-\ell\}$ (using ${\mathcal S}(\rho_A) = {\mathcal S}(\rho_B)$). However, under the assumptions of one-particle localization (\ref{eq:eigcorloc}) and non-degeneracy (\ref{eq:nondeg}) on the XY chain, we have the uniform area law
\begin{equation} \label{statent}
\E \left( \sup_{\rho} {\mathcal E}(\rho) \right) \le C < \infty,
\end{equation}
where the supremum is taken over $\rho = |\psi \rangle \langle \psi|$ for all normalized eigenstates $\psi$ of $H$ and the constant $C$ can be chosen uniformly in $n$ and $\ell$. As before, this includes the disorder average $\E(\cdot)$.

We may also consider the dynamical growth of entanglement under a quantum quench. For this, let $\phi_A$ and $\phi_B$ be eigenstates of $H_A$ and $H_B$, the restrictions of the XY Hamiltonian to the left and right end chains, respectively. Starting with the product state $\rho = |\varphi_A \otimes \varphi_B \rangle \langle \varphi_A \otimes \varphi_B |$, we consider the time evolution $\tau_t(\rho) = e^{itH} \rho e^{-itH}$ under the full XY chain Hamiltonian. Under assumptions as above, the dynamical entanglement remains bounded,
\begin{equation} \label{dynent}
\E \left( \sup_{t, \varphi_A, \varphi_B} {\mathcal E}(\tau_t(\rho)) \right) \le C < \infty,
\end{equation}
uniform in $n$ and $\ell$, with supremum taken over all times $t\in\R$ as well as all eigenstates of $H_A$ and $H_B$. In fact, this result extends to any finite number of quenches in the initial state, with bounds not depending on the number of quenches.

The proofs of (\ref{statent}) and (\ref{dynent}) in \cite{AR-S} and \cite{ANSS} rely heavily on the fact that the eigenstates $\rho$ of $H$ are quasi-free, as already used in the previous section, so that they are entirely determined by their two-point function with respect to the fermionic modes $c_j$, $c_j^*$, summarized in the correlation matrix
\begin{equation} \label{cormatrix}
\Gamma_{\rho} = \rho( {\mathcal C} {\mathcal C}^* ),
\end{equation}
which we use as vector notation for the $2n\times 2n$-matrix with $2\times 2$-matrix-valued entries
\begin{equation}
\Gamma_{\rho}(j,k) = \begin{pmatrix} \tr c_j c_k^* \rho & \tr c_j c_k \rho \\ \tr c_j^* c_k^* \rho & \tr c_j^* c_k \rho \end{pmatrix}, \quad 1\le j, k \le n.
\end{equation}
As observed in \cite{VLRK}, the entropy of quasi-free states can be calculated via the trace identity $\Tr \rho \log \rho = \tr \Gamma_{\rho} \log \Gamma_{\rho}$. Here we distinguish the traces $\tr$ and $\Tr$ in the one-particle space and many-particle space, respectively.
 Most importantly, the reduction $\rho_A$ of $\rho$ to the left end $[1,\ell]$ of the chain is again quasi-free, so that
\begin{equation} \label{traceid}
{\mathcal E}(\rho) = - \Tr \rho_A \log \rho_A = - \tr \Gamma_{\rho}^A \log \Gamma_{\rho}^A.
\end{equation}
Here the correlation matrix $\Gamma_{\rho}^A$ of the reduced state is simply the upper left $2\ell \times 2\ell$-block of $\Gamma_{\rho}$ (by the `left locality' of the operators $c_j$ and $c_j^*$).

By a calculation due to \cite{PasturSlavin} this leads to the bound
\begin{equation} \label{PScalculation}
 {\mathcal E}(\rho) \le 2 \ln 2 \sum_{j=1}^{\ell} \sum_{k=\ell+1}^n \| \Gamma_{\rho}(j,k)\|.
 \end{equation}

The eigenstates of $H$ are given by the fermionic basis vectors $\psi_{\alpha}$ in (\ref{eigenstates}). This allows to calculate the correlation matrices of $\rho=\rho_{\alpha} = |\psi_{\alpha}\rangle \langle \psi_{\alpha}|$ via the Bogoliubov transform ${\mathcal B} = W{\mathcal C}$ and (\ref{diagM}) as
\begin{equation} \label{specproj}
\Gamma_{\rho_{\alpha}} = \chi_{\Delta_{\alpha}}(M).
\end{equation}
Here the right hand side is a spectral projection for the effective Hamiltonian $M$ onto $\Delta_{\alpha} := \{\lambda_j: \alpha_j=0\} \cup \{ -\lambda_j: \alpha_j=1\}$.
Thus (\ref{traceid}) and exponential eigencorrelator localization (\ref{eq:eigcorloc}) lead to
\begin{equation} \label{expbound}
 \E(\sup_{\alpha}{\mathcal E}(\rho_{\alpha})) \le 2\ln 2 \,C \sum_{j=-\infty}^0 \sum_{k=1}^{\infty} e^{-\eta|j-k|} = \frac{2\ln 2\, C e^{-\eta}}{(1-e^{-\eta})^2},
 \end{equation}
proving ({\ref{statent}).

Similar arguments lead to the bound (\ref{dynent}) on dynamical entanglement in \cite{ANSS}. In particular, this uses that the product states $\rho = |\varphi_A \otimes \varphi_B \rangle \langle \varphi_A \otimes \varphi_B |$ as well as their time evolution $\tau_t(\rho)$ under $H$ are still quasi-free, so that ${\mathcal E}(\tau_t(\rho))$ can be calculated using (\ref{traceid}). In this case the correlation matrix is given by
\begin{equation}
\Gamma_{\tau_t(\rho)} = e^{-2itM} \left( \chi_{\Delta_A}(M_A) \oplus \chi_{\Delta_B}(M_B) \right) e^{2itM},
\end{equation}
where $M_A$ and $M_B$ are restrictions of the effective Hamiltonian to $A$ and $B$, respectively, and spectral projections analogous to (\ref{specproj}) appear. Similar to above, now using eigencorrelator localization for $M$, $M_A$ and $M_B$, this leads to (\ref{dynent}).

Several remarks are in order here:

(i) The bound (\ref{expbound}) on the entanglement grows as $O(\xi^2)$ as the one-particle localization length $\xi = 1/\eta$ diverges. This is substantially stronger than what one gets by combining the general result of \cite{BrandHor15} with the bound on correlation decay from \cite{SimsWarzel}, giving the bound $O(e^{c \xi \log \xi})$ on the entanglement.

(ii) The argument leading to (\ref{PScalculation}) was previously used in \cite{PasturSlavin} to bound the entanglement in disordered $d$-dimensional free Fermion systems, where the general area law $O(\ell^{d-1})$ is found, both as upper and lower bound. Moreover, it was shown in \cite{EPS} that the fermionic entanglement is self-averaging for $d>1$ but not for $d=1$. Note, however, that the fermionic entanglement has to be interpreted differently from entanglement in spin systems due to the non-locality of the fermionic modes, e.g.\ \cite{Banulsetal, PeschelEisler}, and that the connection to the XY chain via Jordan-Wigner only holds for $d=1$.

(iii) In principle, the above result also applies to thermal states $\rho_{\beta} = e^{-\beta H}/ \tr e^{-\beta H}$. In this case, instead of (\ref{specproj}) the correlation matrix is $\Gamma_{\rho_{\beta}} = (\idty + e^{-2\beta M})^{-1}$, so that the eigencorrelator bound (\ref{eq:eigcorloc}) still applies and gives exponential decay of matrix elements and an area law for the bipartite entanglement ${\mathcal S}(\rho_{\beta})$. However, for the mixed state $\rho_{\beta}$ this is generally not considered a good entanglement measure (in particular, vanishing of ${\mathcal S}(\rho_{\beta})$ does not mean that $\rho_{\beta}$ is a product state). A better quantity to consider is the logarithmic negativity of $\rho_{\beta}$, e.g.\ \cite{VidalWerner, Plenio}, but it remains an open problem if the logarithmic negativity of mixed states of the disordered XY chain satisfies an area law.

On the other hand, it follows easily from (\ref{statent}) that the {\it entanglement of formation} of thermal states satisfies an area law. The latter is the convex roof extension of the entanglement entropy \cite{Wootters}, defined for a general mixed state $\rho$ as
\begin{equation}
E_f(\rho) = \min_{p_k, \phi_k} \sum_k p_k {\mathcal E}(|\phi_k\rangle \langle \phi_k|),
\end{equation}
with arbitrary $p_k$ and unit vectors $\phi_k$ such that $\rho = \sum_k p_k |\phi_k\rangle \langle \phi_k|$.

The thermal states are given by \[\rho_{\beta} = Z_{\beta}^{-1} \sum_{\alpha} e^{-\beta q_{\alpha}} |\psi_{\alpha}\rangle \langle \psi_{\alpha}|,\] where $\psi_{\alpha}$ and $q_{\alpha}$ are the eigenstates and eigenvalues of $H$ and $Z_{\beta} = \sum_{\alpha} e^{-\beta q_{\alpha}}$. Using the uniform area law (\ref{statent}) for the eigenstates $\psi_{\alpha}$ this immediately leads to
\begin{equation} \label{entform}
\E \left( \sup_{\beta} E_f(\rho_{\beta}) \right) \le C < \infty,
\end{equation}
an area law for the averaged entanglement of formation of the thermal states $\rho_{\beta}$, uniform in the inverse temperature $\beta$.

One also gets a similar extension of (\ref{dynent}) to the case where $\rho= \rho_A \otimes \rho_B$ and $\rho_A$ and $\rho_B$ are thermal states of the subsystems.

\section{Absence of particle and energy transport} \label{sec:parttrans}

Here we consider the isotropic XY chain, i.e.\ we set $\gamma_j=0$ for all $j$ in (\ref{XYchain}). Thus the matrix $B$ in (\ref{B}) vansihes and, using the CAR in (\ref{JWrep}), the Hamiltonian can be written as
\begin{eqnarray} \label{Hiso}
H_{iso} & = & - \sum_{j=1}^{n-1} \mu_j (\sigma_j^X \sigma_{j+1}^X + \sigma_j^Y \sigma_{j+1}^Y) - \sum_{j=1}^n \nu_j \sigma_j^Z \nonumber \\ & = & 2\sum_{j,k} c_j^* A_{jk} c_k +\tilde{E}_0 \idty,
\end{eqnarray}
where $\tilde{E}_0 = \sum_j \nu_j$. This simpler form of the quadratic Fermionic Hamiltonian (as compared to (\ref{JWrep})) allows diagonalization via $(\tilde{b}_1,\ldots,\tilde{b}_n)^t := U (c_1,\ldots, c_n)^t$, i.e.\ a Bogoliubov transformation which does not mix creation and annihilation modes. Here $U$ is the orthogonal matrix which diagonalizes $A$:
\begin{equation} \label{diagA}
UAU^t = \mbox{diag}(\tilde{\lambda}_j),
\end{equation}
with $\tilde{\lambda}_j$ the eigenvalues of $A$. One gets the free Fermion system
\begin{eqnarray} \label{freeFermion}
H_{iso} = 2\sum_j \tilde{\lambda}_j \tilde{b}_j^* \tilde{b}_j + \tilde{E}_0 \idty.
\end{eqnarray}
Let $\psi_0$ be the vacuum vector associated with this free Fermion system and
\begin{equation} \label{fockbasis}
\psi_{\beta} = \prod_{j=1}^n (\tilde{b}_j^*)^{\beta_j} \psi_0, \quad \beta \in \{0,1\}^n,
\end{equation}
the corresponding Fermionic basis of eigenvectors of $H_{iso}$. Due to the non-mixing of modes, $\psi_0$ coincides with the all spins down vector $e_0$ and, for each $0\le k \le n$, the $k$-particle space span$\{\psi_{\beta}: \#\{j:\beta_j=1\}=k\}$ coincides with the space spanned by the spin basis vectors $e_{\alpha}$ with $k$ up-spins. Thus the number of up-spins is interpreted as the number of particles and ${\mathcal N} = \sum_{j=1}^n a_j^*a_j = \sum_{j=1}^n c_j^* c_j$ as the total particle number operator. Preservation of the particle number if reflected in the fact that $H_{iso}$ commutes with ${\mathcal N}$.

For subsets $S\subset [1,n]$ the local particle number operators are ${\mathcal N}_S = \sum _{j\in S} a_j^* a_j$. For any state $\rho$ the expected number of particles (up-spins) in $S$ is given by $\langle {\mathcal N} \rangle_{\rho} = \Tr \rho {\mathcal N}_S$.

In the many-body localized phase one expects absence of particle transport. In fact, let $S_1$ and $S_2$ be two disjoint subsets of $\Lambda$ such that $S_2\subset\Lambda\setminus[\min S_1,\max S_1]$, i.e., $S_1$ is entirely surrounded by $S_2$. As initial state consider a particle profile
\begin{equation} \label{profile}
 \rho = \bigotimes_{j=1}^n \rho_j, \quad \rho_j = \begin{pmatrix} \eta_j & 0 \\ 0 & 1-\eta_j \end{pmatrix},
 \end{equation}
and assume $\eta_j=0$ for all $j\in \Lambda \setminus S_2$, meaning only down-spins outside $S_2$.  Let $\rho_t = e^{-iHt} \rho e^{iHt}$ be the Schr\"odinger evolution. If exponential eigencorrelator localization (\ref{eq:eigcorloc}) is assumed, then it  can be shown that
\begin{equation} \label{particletrans}
\E \left( \sup_t \langle {\mathcal N}_{S_1} \rangle_{\rho_t} \right) \le \frac{2C}{(1-e^{-\eta})^2} e^{-\eta d(S_1,S_2)}.
\end{equation}
Here $d(S_1,S_2) = \min\{|x-y|: x \in S_1, y\in S_2\}$ is the distance of $S_1$ and $S_2$. This bound follows as a special case of Theorem~1.1 in \cite{ANSS}. The argument is similar to the proof of Theorem~\ref{thm:energytrans} below, provided in Appendix~\ref{appA}.

Thus, if at time $t=0$ all particles are concentrated in $S_2$ (corresponding to $\eta_j=1$ for $j\in S_2$), then the number of particles at distance more than $d$ from $S_2$ remains exponentially small in $d$, uniformly for all times $t$. As in the previous section, the bound diverges as $O(\xi^2)$ in the one-particle localization length $\xi=1/\eta$. The proof of (\ref{particletrans}) can be seen from the proofs of Theorems \ref{thm:energytrans} and \ref{thm:energytrans2} below.

Additional results of this type, establishing the absence of particle transport in the disordered isotropic XY chain, can be found in \cite{ANSS}. Similar results for the disordered Tonks-Girardeau gas, which can be viewed as a continuum analogue of the XY chain, can be found in \cite{SeiringerWarzel}.

This result does not extend to the anisotropic XY chain, were the particle number is not preserved and thus particles can be created by local properties of the dynamics, even if the system if fully in the MBL phase.

In essentially the same way one can show the absence of energy transport in the MBL phase. For this let $S_1$, $S_2$ be as above, but assume that $S_1=[a,b]$ is an interval. By $H_{S_1}$ we denote the restriction of the isotropic XY chain to $S_1$, given as in (\ref{Hiso}) with $[1,n]$ replaced by $[a,b]$ (but still acting on the full spin chain). Thus $\langle H_{S_1} \rangle_{\rho}$ is the expected energy of a state $\rho$ in $S_1$.

\begin{thm} \label{thm:energytrans}
Consider the isotropic XY chain $H=H_{iso}$ as in (\ref{Hiso}) and the particle profile $\rho$ as in (\ref{profile}), with $\eta_j=0$ for all $j\in\Lambda\setminus S_2$. Then under the assumption of eigencorrelator localization (\ref{eq:eigcorloc}) we have
\begin{equation} \label{energytrans}
\E \left( \sup_t | \langle H_{S_1} \rangle_{\rho_t} - \tilde{E}_0| \right) \le \frac{4CD}{(1-e^{-\eta})^2} e^{-\eta d(S_1,S_2)}.
\end{equation}
Here $\tilde{E}_0 = \langle H_{S_1} \rangle_{\rho} = \sum_{j\in S_1} \nu_j$ and $D$ is a uniform upper bound on the matrix norm of the Anderson model $A$ in (\ref{A}).
\end{thm}
The uniform matrix bound $D$ can be chosen to be for example $D=2\mu_{max}+\nu_{max}$, where $\mu_{max}$ and $\nu_{max}$ are the maximal values of the distributions of $|\mu_j|$ and $|\nu_j|$, respectively. To interpret (\ref{energytrans}), note that by construction of $\rho$ at time $t=0$ all its interaction energy is concentrated in $S_2$, so that the amount of energy which can be transported from $S_2$ to $S_1$ is exponentially small in the distance of the two subsystems, uniformly in time.


A weakened version of (\ref{thm:energytrans}) can be shown to hold for the more general anisotropic XY chain (\ref{XYchain}).
\begin{thm}\label{thm:energytrans2}
Consider the anisotropic XY chain $H$ given in (\ref{XYchain}) with the assumption of eigencorrelator localization (\ref{eq:eigcorloc}). Let $\rho$ be the general particle profile as in (\ref{profile}). Then there is a constant $C'<\infty$, depending on the parameters of the Hamiltonian, but independent of the sizes of $\Lambda$ and $S_1$, such that
\begin{equation} \label{energytrans2}
\E \left( \sup_t |\langle H_{S_1} \rangle_{\rho_t} - \langle H_{S_1} \rangle_{\rho} | \right) \le C',
\end{equation}
\end{thm}
The constant bound in (\ref{energytrans2}) means that the energy fluctuations within the subsystem $S_1$ are uniformly bounded from above, while the full energy $\langle H \rangle_{\rho}$ will grow linearly in the size of $\Lambda$.

Detailed proofs of Theorems~\ref{thm:energytrans} and \ref{thm:energytrans2} are provided in Appendix~\ref{appA}.


\section{Fock space localization} \label{sec:fock}

Many-body localization of eigenstates is generally expected to correspond to localization in Fock space, not in physical space, e.g.\ \cite{Altshuleretal, Baskoetal, BauerNayak}. This refers to a picture where the Fock space basis of Slater determinants of the (non-interacting) single-particle eigenstates is used as a reference lattice, and eigenstates of the interacting system can be labelled by the Fock space basis and decay rapidly (exponentially) in the distance from some finite subset of the lattice. Here we demonstrate this form of MBL for the disordered XY chain.

As in Section~\ref{sec:parttrans} we consider the isotropic XY quantum spin chain in transversal field
\begin{equation} \label{XYlargedisorder}
H_{\varepsilon}= -\varepsilon \sum_{j=1}^{n-1} (\sigma_j^X \sigma_{j+1}^X + \sigma_j^Y \sigma_{j+1}^Y) - \sum_{j=1}^n \nu_j \sigma_j^Z = 2 c^* A c + \tilde{E}_0,
\end{equation}
but now for the case of high disorder, meaning we choose the constant $\varepsilon>0$ close to zero. Thus the effective Hamiltonian
\begin{equation}
A = \left( \begin{array}{cccc} -\nu_1 & \varepsilon & & \\ \varepsilon & \ddots & \ddots & \\ & \ddots & \ddots & \varepsilon \\ & & \varepsilon & -\nu_n \end{array} \right)
\end{equation}
is the Anderson model on $[1,\ldots,n]$ at high disorder.

Elgart and Klein \cite{EK} have recently provided a variant of the multiscale analysis method, based on a direct analysis of eigenvectors and eigenvalues rather than on decay properties of the Green function, which allows to show that, with high probability, the eigenvectors of the Anderson model at high disorder (in any dimension) are close to the canonical basis vectors, uniformly in all lattice sites. In fact, this property can also be derived from eigencorrelator localization (\ref{eq:eigcorloc}), which we have used throughout this work as a unifying criterion. Let us describe this connection.

Under the assumptions in Section~\ref{sec:onepart}, i.e.\ for example if the distribution of the $\nu_j$ has a bounded density of compact support, one can show the existence of $C<\infty$ and $\mu>0$, such that
\begin{equation} \label{ecorloceps}
\E\left(\sup_{|g|\le 1} |g(A)_{jk}|\right) \le C e^{-\mu |\log \varepsilon| |j-k|}
\end{equation}
for all $n\in \N$, $\varepsilon>0$ and $j,k \in [1,\ldots,n]$. This follows, e.g., from bounds provided in Chapters 6 and 7 of \cite{AW}  for the equivalent case of large coupling $\lambda=1/\varepsilon$.

\begin{lem} \label{lem:etaloc}
Let $0<\tau<1$ and $\eta>0$ be given. Then there exist $\varepsilon_0 >0$ and $\xi>0$ such that, for $n\in \N$ and all $0<\varepsilon \le \varepsilon_0$, the Anderson model $A$ with probability at least $1-e^{-n^\xi}$ has  an orthonormal basis of eigenfunctions $\varphi_k$, $k\in {1,\ldots,n}$, such that
\begin{equation} \label{expdecay}
|\varphi_k(j)| \le e^{-\eta|k-j|} \quad \mbox{for all $j, k \in \{1,\ldots,n\}$ with $|j-k| \ge n^{\tau}$}.
\end{equation}
\end{lem}

We will show in Appendix~\ref{proof8} how this follows from (\ref{ecorloceps}). It also follows from Theorem~1.6 in \cite{EK} where somewhat weaker assumptions on the distribution of the $\nu_j$ are required. While we only state the result for the one-dimensional Anderson model, the argument extends to arbitrary dimension.

We now describe how this effects the many-body eigenvectors. A basis of eigenvectors of the non-interacting system $H_{\varepsilon=0}$ is given by the spin basis $e_{\alpha}$, $\alpha \in \{0,1\}^n$. We re-label this basis by Fermionic particle configurations, i.e.\ by the lattice sites of the antisymmetric Fock space,
\begin{equation} \label{Focklabels}
\{ (j_1,\ldots,j_r): 0\le r \le n, 1\le j_1< \ldots < j_r \le n\}
\end{equation}
as $\tilde{e}_{(j_1,\ldots,j_r)} = e_{\alpha}$ with $\alpha\in \{0,1\}^n$ such that $\{j: \alpha_j=1\} = \{j_1<\ldots<j_r\}$. These correspond to the Slater determinants of the eigenvectors $e_j$, $j=1,\ldots,n$, of the effective Hamiltonian $A$ (at $\varepsilon=0$).

Similarly, we label the eigenvectors (\ref{fockbasis}) of the interacting system by particle configurations with respect to the free Fermion system (\ref{freeFermion}) as $\tilde{\psi}_{(k_1,\ldots,k_r)} = \psi_{\beta}$, $\beta\in \{0,1\}^n$ such that $\{k:\beta_k=1\} = \{k_1,\ldots,k_r\}$. In this basis $H_{iso}$ is represented as $2d\Gamma_a(A) + \tilde{E}_0 \idty$, where $d\Gamma_a(A)$ is the restriction of the second quantization of the effective Hamiltonian $A$ to the anti-symmetric Fock space ${\mathcal F}_a(\C^n)$.

Fock space localization of $H_{\varepsilon}$ means that the Fourier coefficients $\langle \tilde{\psi}_{(k_1,\ldots,k_r)}, \tilde{e}_{(j_1,\ldots,j_r)} \rangle$ of its eigenvectors with respect to the basis of the non-interacting system decay rapidly in the distance of the Fermionic configurations $k=(k_1,\ldots,k_r)$ and $j=(j_1,\ldots,j_r)$. As in Section~\ref{sec: correlations} we choose the latter as $D(j,k) := \max_{\ell \in \{1,\ldots,r\}} |j_{\ell} - k_{\ell}|$.
The first part of the following result establishes Fock space localization of $H_{\epsilon}$ with respect to this distance.

The second part is a result of similar flavor with a more direct proof. Here we use the local spin occupation operators $n_x := c_x^* c_x = a_x^* a_x$ to express closeness of the vectors $\tilde{\psi}_k$ to the non-interacting basis vectors.

\begin{thm} \label{thm:fock}
For given $0<\tau<1$ and $\eta>0$, let $\varepsilon_0$ and $\xi$ be as above and choose $\eta_0<\eta$. Then, for all $n \in \N$ and $0<\varepsilon \le \varepsilon_0$, it holds with probability at least $1-e^{-n^{\xi}}$ that:

(a) There is a constant $C=C(\eta,\tau)$ such that for every $1\le r \le n$ and every pair of Fermionic configurations $k=(k_1,\ldots,k_r)$ and $j=(j_1,\ldots,j_r)$ with $D(j,k) \ge 2n^{\tau}$,
\begin{equation} \label{Fockloc}
\left| \langle \tilde{\psi}_{k}, \tilde{e}_{j} \rangle \right|  \le C n^{2\tau} e^{-\frac{\eta-\eta_0}{4} D(j,k)}.
\end{equation}

(b) For every $1\le x \le n$ and every Fermionic configuration $k=(k_1,\ldots,k_r)$ with $\min_{\ell} \{|k_{\ell}-x|\} \ge n^{\tau}$,
\begin{equation} \label{occnumberbound}
\langle \tilde{\psi}_{k}, n_x \tilde{\psi}_{k} \rangle  \le \frac{2}{e^{2\eta \min_{\ell} \{|k_{\ell}-x|\}}-1}.
\end{equation}

\end{thm}

These results have not  been previously published and we provide detailed proofs in Appendix~\ref{proof8}.

\begin{remark} (i) Note that $n_x$ is the orthogonal projection onto the space spanned by the spin product states $e_{\alpha}$ with $\alpha_x=1$, so that $\langle \tilde{\psi}_{k}, n_x \tilde{\psi}_{k} \rangle$ is the component of $\tilde{\psi}_k$ in that space. (\ref{occnumberbound}) says that this component is small if the configuration $k$ contains no ``particles'' near $x$.

(ii) The need for the condition $|j-k|\ge n^{\tau}$ is not purely
an artifact of the method of proof. Since in a disordered system localization is {\em not}, in general, uniform, a condition of this form is not entirely avoidable. For example, as the random potential will be essentially constant on arbitrarily long intervals with high probability, there will be eigenfunctions which are widely spread out before transitioning to exponential decay as in (\ref{expdecay}). An informative account of this issue can be found in \cite{delrio:1995}. Note, however, that $\tau>0$ can be chosen arbitrarily small. As a consequence of the proof this yields that $\xi$ is close to zero as well. An interesting open question is if disorder averaging might eliminate the restriction $|j-k|\ge n^{\tau}$ from (\ref{expdecay}), i.e.\ if exponential decay of $\E(|\varphi_k(j)|)$ can be proven for the Anderson model at large disorder.
\end{remark}

\section{Summary and Outlook}

In this paper we reviewed several aspects of MBL in the context of the disordered XY chains: localization properties of the energy eigenstates
and thermal states, propagation bounds of Lieb-Robinson type, decay of correlation functions, absence of transport of particles and energy, bounds on bipartite entanglement, the dynamical generation of entanglement, eigenstate labeling and so-called Fock space localization.
The exact mapping of the XY chain to a system of quasi-free fermions and access to detailed information about the localization properties of the related one-body operator, play an essential role in the derivation of these results for the XY chain. We believe that having detailed, rigorous, and quite specific results for the XY chain is helpful in shaping our understanding of the subtle issues raised by MBL, and can be used as a stepping stone to more general mathematical results about MBL in disordered quantum spin systems and other quantum many-body systems.

We conclude with a discussion of two conjectures and further directions for mathematical research on MBL. The first conjecture is
 concerned with one-dimensional systems of spins or fermions. Based on numerical work on the disordered quantum Ising chain in a random field, it was conjectured that such a one-dimensional system,
at sufficiently large disorder exhibits full MBL in the sense that there exists a (random) quasi-local unitary transformation $U(\omega)$ that maps the Hamiltonian $H(\omega)$ in to a random but diagonal Ising-type model:
\begin{equation}
U(\omega)^* H(\omega) U(\omega) = \sum_{X}  K_X(\omega) \sigma_X^Z,
\end{equation}
where the sum runs over finite subsets $X=\{j_1,\cdots,j_k\}$ of the one-dimensional lattice, $\sigma^Z_X
=\sigma^Z_{j_1}\cdots \sigma^Z_{j_k}$. The random couplings are short-range, e.g., in the sense that there are constants
$a, M$ such that
\begin{equation}
\sup_{j,k} e^{a | j-k|} \sum_{\substack{X\\ j,k\in X}} | K_X(\omega) |\leq M,
\end{equation}
in a suitable probabilistic sense. A good quasi-locality property of $U(\omega)$ would be the following Lieb-Robinson-type inequality with a sufficiently fast decaying
function $F$ and an exponent $p\geq 0$: for any pair of observables $A$ and $B$ supported on finite sets $X$ and $Y$, one has
\begin{equation}
\Scale[0.9]{\displaystyle\Vert [ U(\omega)^* A U(\omega), B]\Vert  \leq \Vert A\Vert\, \Vert B\Vert |X|^p F(\min \{ |j-k| \mid j\in X, k\in Y\}).}
\end{equation}
The work by Imbrie on the random quantum Ising chain provides evidence for such a property \cite{imbrie:2014}, but requires an
unproven assumption of level repulsion in the spectrum of the system. Ideally, one would
want a general result that applies to any sufficiently disordered one-dimensional lattice system with finite-range interactions, without
hard to verify assumptions about their eigenvalue statistics.

The second conjecture generalizes the zero-velocity Lieb-Robinson bounds found for the XY chain to general disordered finite range
spin models on $\mathbb{Z}^d$. For $d\geq 2$, MBL may be seen only at low energies, although in what sense is still a matter of debate
(see, e.g., \cite{deroeck:2014, deroeck:2015}).

Let $P_E$ denote the spectral projection onto the energy interval $[E_0,E_0+E]$, with $E_0$ the ground state energy.
We would then like to have constants $C, c, \mu >0$, and $q\geq 0$, a function $g_E(t)$ such that for any pair of finite disjoint
subsets $X,Y\subset \mathbb{Z}^d$, and observables $A$ and $B$  supported on $X$ and $Y$, respectively, one has for
$0\leq E\leq c {\rm dist}(X,Y)^q$,  the bound
\begin{equation}
\mathbb{E} (\Vert [\tau_t^\omega(P_E A P_E), B]\Vert)
\leq C \Vert A\Vert \Vert B\Vert \vert X\vert  g_E(t) e^{-\mu d(X,Y)}, t\in \mathbb{R}.
\end{equation}
Here, $\mathbb{E}$ denotes the expectation with respect to the randomness, $\tau_t^\omega$ is the Heisenberg dynamics
generated by the random Hamiltonian, and one is interested
in showing the inequality with a function $g_E(t)$ of moderate growth. E.g., a good example would be $g_E(t) \sim |t|^\alpha$,
for some $\alpha\geq 0$. The standard Lieb-Robinson bound for systems with a bounded finite-range interaction holds with
$g_E(t)=C \exp(v|t|)$, where $v>0$ is a bound for the Lieb-Robinson velocity and for translation invariant systems ballistic propagation,
i.e., with a positive velocity, is in general expected. So, any function $g_E$ which grows strictly slower than
exponential could provide non-trivial new information.

Among the more refined mathematical properties worth investigating we mention the disorder-averaged distribution of the difference
between consecutive eigenvalues of the many-body Hamiltonian. Numerical study of a simple one-dimensional model indicates a clear
distinction between a delocalized regime at weak disorder and a localized regime at strong disorder \cite{OganesyanHuse}. In the first
the distribution very much looks like the one found for a GOE random matrix, whereas at strong disorder the distribution is consistent
with a Poisson distribution of eigenvalues at not too small values of the spacing. Establishing the existence of these two regimes, and
perhaps also the existence of an intermediate phase, in a bona-fide many body model is certainly a significant mathematical challenge,
but it is only by tackling challenging questions that we have a chance to make progress. In this spirit, it makes sense to also ask about MBL
for bosons, such as phonons in disordered crystal lattices \cite{MonthusGarel}. One could again start by considering disordered models reducible to a one-body operator such as the ones studied in \cite{NSS1,NSS2} and then continue with perturbations of such models.

\section*{Acknowledgements}

The work of BN was supported in part by the National Science Foundation under Grant DMS-1515850. GS acknowledges hospitality at the Steklov Institute of the Russian Academy of Sciences in St.\ Petersburg where part of this work was done.

\bigskip

\appendix

\section{Proof of Theorems~\ref{thm:energytrans} and \ref{thm:energytrans2}} \label{appA}

For the proof of Theorem~\ref{thm:energytrans},
we first write $H_{S_1}=2 c^* A_{S_1}c +\tilde{E}_0\idty$, where $A_{S_1}$ is the restriction of Anderson model $A$ in (\ref{A}) to $S_1$, i.e. $
A_{S_1}:=\chi_{S_1}A\chi_{S_1}$ and $c=(c_1,c_2,\ldots,c_n)^t$. Then
\begin{equation}\label{eq:iso:initial}
\langle H_{S_1}\rangle_{\rho}=\sum_{j,k=1}^n(A_{S_1})_{jk}\langle c_j^*c_k\rangle_{\rho}+\tilde{E}_0= \tilde{E}_0
\end{equation}
where we used that $\langle c_j^*c_k\rangle_{\rho}=0$ for $j,k\in S_1$.
For positive times, one can see that
\begin{eqnarray}\label{EnTr:eq:mainstep}
\langle H_{S_1}\rangle_{{\rho}_t}-\tilde{E}_0 & = & \langle\tau_t( H_{S_1})\rangle_{\rho}-\tilde{E}_0=\langle 2\tau_t(c^*)A_{S_1} \tau_t(c)\rangle_{{\rho}} \nonumber \\
& = & 2 \sum_{j,k=1}^n (e^{2itA} A_{S_1} e^{-2itA})_{jk} \langle c^*_jc_k\rangle_\rho
\end{eqnarray}
where we used $\tau_t(c)=e^{-2itA} c$, the isotropic case of (\ref{dynrelation}). The sum in (\ref{EnTr:eq:mainstep}) can be written as a trace,
\begin{equation}\label{EnTr:eq:tr}
\langle H_{S_1}\rangle_{\rho_t}-\tilde{E}_0= 2\tr e^{2itA} A_{S_1}e^{-2itA}\eta \chi_{S_2}
\end{equation}
where $\eta=\eta\chi_{S_2}$ is the diagonal $n\times n$ matrix supported on $S_2$ with diagonal elements $1-\eta_j$ for $j\in S_2$. This means that
\begin{eqnarray}
\left|\langle \tau_t(H_{S_1})- \tilde{E}_0\rangle_{\rho_t}\right|&\leq& 2\left|\tr \chi_{S_2} e^{2itA} \chi_{S_1}A\chi_{S_1}e^{-2itA}\eta\right| \\
& \leq & 2\| \chi_{S_2} e^{2itA} \chi_{S_1} \|_1 \| A\chi_{S_1}e^{-2itA}\eta \| \nonumber \\
& \leq & 2\|A\| \sum_{j\in S_2} \sum_{k\in S_1} |(e^{2itA})_{jk}|, \nonumber
\end{eqnarray}
where we used cyclicity of the trace in the first step and that the trace norm can be bounded by the sum of the absolute values of the matrix elements.

Bound (\ref{energytrans}) follows by taking the sup over all times, then averaging and using eigencorrelator localization (\ref{eq:eigcorloc}).

In the anisotropic case of Theorem~\ref{thm:energytrans2}, corresponding to (\ref{quadform}) we have
\begin{equation}
H_{S_1}=(\mathcal{C})^* M_{S_1}\mathcal{C},
\end{equation}
where $\mathcal{C}=(c_1,c_1^*,\ldots,c_n,c_n^*)^t$ and $M_{S_1}$ is the $2\times 2-$block restriction of the matrix $M$ given in (\ref{effHamanisotropic}) to $S_1$, i.e., the restriction to $\Span\{\delta_{2j-1},\delta_{2j},\ j\in S_1\}$,
\begin{equation}
M_{S_1}:=\chi_{S_1}M\chi_{S_1}.
\end{equation}

One can easily see that
\begin{eqnarray}\label{eq:ave:initial}
\langle H_{S_1}\rangle_\rho=\langle \mathcal{C}^*M_{S_1} \mathcal{C}\rangle_{\rho} & = &
\sum_{j,k=1}^{2 n} \langle \delta_j, M_{S_1}\delta_k\rangle\langle(\mathcal{C}^*)_j(\mathcal{C})_k\rangle_{\rho} \nonumber \\ & = & \tr \chi_{S_1}\Gamma M,
\end{eqnarray}
with the correlation matrix $\Gamma=\diag\{\eta_1,1-\eta_1, \ldots, \eta_n,1-\eta_n\}$. Here we recall that $\chi_{S_1}$ is the $2\times 2$-block restriction to $S_1$, we also need to stress here that we used $\langle \delta_j, M_{S_1}\delta_k\rangle$ to refer to the $jk$-th element (not block) of $M_{S_1}$. Likewise, $(\mathcal{C}^*)_j$ is the $j$-component of the vector $\mathcal{C}^*$.

By evolving the system in time, one can see that
\begin{eqnarray}\label{EnTr:eq:Aniso:tr}
\langle H_{S_1}\rangle_{\rho_t} & =& \langle \tau_t(\mathcal{C}^*)M_{S_1} \tau_t(\mathcal{C})\rangle_{\rho}\nonumber\\
& = &  \langle \mathcal{C}^* e^{2itM} M_{S_1} e^{-2itM}\mathcal{C}\rangle_\rho \nonumber \\ & = & \tr e^{2itM}M_{S_1} e^{-2itM}\Gamma
\end{eqnarray}
where we used (\ref{dynrelation}).

Note that (\ref{EnTr:eq:Aniso:tr}) can be written as
\begin{eqnarray}\label{eq:HS}
\langle H_{S_1}\rangle_{\rho_t} & = & \tr \chi_{S_1} e^{-2itM}\Gamma \chi_{S_1} e^{2itM}\chi_{S_1} M \nonumber \\ & \mbox{ } & \quad  + \tr \chi_{S_1} e^{-2itM}\Gamma \chi_{S_1^c} e^{2itM}\chi_{S_1} M
\end{eqnarray}
And since we are looking at the difference $\langle\tau_t(H_{S_1})-H_{S_1}\rangle_\rho$, we can rewrite (\ref{eq:ave:initial}) to match (\ref{eq:HS}) as follows
\begin{eqnarray}\label{eq:HSt}
\langle H_{S_1}\rangle_{\rho}
& = & \tr \chi_{S_1} e^{-2itM} \Gamma \chi_{S_1} e^{2itM} \chi_{S_1} M \nonumber \\ & \mbox{ } & + \tr \chi_{S_1} e^{-2itM}\chi_{S_1} \Gamma e^{2itM} \chi_{S_1^c} M
\nonumber \\ & \mbox{ } & \quad +\tr \chi_{S_1^c} e^{-2itM}\chi_{S_1} \Gamma M e^{2itM}
\end{eqnarray}
Then subtracting (\ref{eq:HSt}) from (\ref{eq:HS}) we get
\begin{eqnarray}
\langle\tau_t(H_{S_1})-H_{S_1}\rangle_\rho & = & \tr \chi_{S_1} e^{-2itM}  \chi_{S_1^c}\Gamma  e^{2itM}\chi_{S_1} M \nonumber \\
& \mbox{ } & \quad -\tr \chi_{S_1} e^{2itM} \chi_{S_1^c} M\chi_{S_1} e^{-2itM} \Gamma \nonumber \\
&\mbox{ } & \quad \quad -\tr \chi_{S_1^c} e^{-2itM}\chi_{S_1} \Gamma M e^{2itM} \nonumber
\end{eqnarray}
where we used $[M,e^{2itM}]=[\Gamma,\chi_{S_1}]=0$. By taking the absolute value
\begin{eqnarray}
\left|\langle\tau_t(H_{S_1})-H_{S_1}\rangle_\rho\right| &\leq& \|\chi_{S_1} e^{-2itM}  \chi_{S_1^c}\|_1 \|\Gamma e^{2itM}\chi_{S_1} M\| \\ & & + \|\chi_{S_1} e^{2itM}  \chi_{S_1^c}\|_1 \|M \chi_{S_1} e^{-2itM}\Gamma\| \nonumber \\ & & + \|\chi_{S_1^c} e^{-2itM}  \chi_{S_1}\|_1 \|\Gamma M e^{2itM}\| \nonumber\\
&\leq& 3\cdot 2\|M\| \sum_{j\in S_1} \sum_{k\in S_1^c} \|(e^{2itM})_{jk}\|. \nonumber
\end{eqnarray}
Here $(e^{2itM})_{jk}$ are the $2\times 2$-block elements of $e^{2itM}$ and we used the simple fact that for any $2n\times 2n$ matrix $A$ with $2\times 2$ block elements $A_{jk}$ we have $\|A\|_1\leq 2\sum_{j,k=1}^n \|A_{jk}\|$.

The bound (\ref{energytrans2}) follows by taking the supremum over all times, then averaging and using eigencorrelator localization (\ref{eq:eigcorloc}).


\section{Proof of Lemma~\ref{lem:etaloc} and Theorem~\ref{thm:fock}} \label{proof8}

Towards the proof of Lemma~\ref{lem:etaloc} we start by noting that the spectrum of $A$ is almost surely non-degenerate. Thus we may label the eigenvalues of $A$ by $E_1<E_2<\ldots<E_n$ and corresponding normalized eigenfunctions by $\varphi_{E_r}$, $r=1,\ldots,n$. The eigencorrelators may thus be expressed as \cite{AW}
\begin{equation} \label{efcor2}
Q(j,k) =  \sup_{|g|\le 1} |g(A)_{jk}| = \sum_{E\in \sigma(A)} |\varphi_E(j) \varphi_E(k)|.
\end{equation}
By (\ref{ecorloceps}) and Chebychev's inequality we have, for each fixed $j$ and $k$,
\begin{equation} \label{cheby}
\mathbb{P}\left(Q(j,k) < e^{-\mu |\log \varepsilon| |j-k|/2} \right) \ge 1 - Ce^{-\mu |\log \varepsilon| |j-k|/2}.
\end{equation}
Fix $\alpha>1$ and for each $r\in [1,\ldots,n]$ define the (random) set
\begin{equation}
\mathcal{N}_r := \left\{j\in [1,\ldots,n]: |\varphi_{E_r}(j)| \ge n^{-\alpha} \right\}.
\end{equation}
Note that due to normalization of $\varphi_{E_r}$ the sets $\mathcal{N}_r$ are non-empty (for this $\alpha \ge 1/2$ would suffice). Fix $k_r \in \mathcal{N}_r$ for each $r$, which will serve as localization centers.

By (\ref{efcor2}) and (\ref{cheby}) we have for every fixed $j$ and $r$ that
\begin{equation} \label{event1}
\mathbb{P}\left( |\varphi_{E_r}(j)| < n^{\alpha} e^{-\mu |\log \varepsilon| |j-k_r|/2} \right) \ge 1 - Ce^{-\mu |\log \varepsilon| |j-k_r|/2}.
\end{equation}
Fix $0<\tau<1$ and let $A_r$ denote the event that $|\varphi_{E_r}(j)| < e^{-\mu |\log \varepsilon| |j-k_r|/4}$ for all $j$ such that $|j-k_r|\ge n^{\tau}$. Also, let $B_r$ be the event that (\ref{event1}) holds simultaneously for all $j$ with $|j-k_r|\ge n^{\tau}$. Then
\begin{eqnarray} \label{Br}
\mathbb{P}(B_r) & \ge & 1- C \sum_{j:|j-k_r| \ge n^{\tau}} e^{-\mu|\log \varepsilon| |j-k_r|/2} \nonumber \\
& \ge & 1- C_1 e^{-\mu |\log \varepsilon| n^{\tau}/2}.
\end{eqnarray}
Assume that $B_r$ holds, then $|j-k_r| \ge (|j-k_r|+ n^{\tau})/2$ implies that
\begin{equation}
|\varphi_{E_r}(j)| < n^{\alpha} e^{-\mu|\log \varepsilon| n^{\tau}/4} e^{-\mu |\log \varepsilon| |j-k_r|/4}
\end{equation}
for all $j$ such that $|j-k_r| \ge n^{\tau}$. For $0<\varepsilon \le \varepsilon_0$ sufficiently small the term $n^{\alpha} e^{-\mu|\log \varepsilon| n^{\tau}/4}$ is uniformly bounded by $1$, so that $B_r \subset A_r$. Thus (\ref{Br}) gives
\begin{equation}
\mathbb{P}(A_r) \ge 1 - C_1 e^{-\mu |\log \varepsilon| n^{\tau}/2}
\end{equation}
and
\begin{equation}
\mathbb{P}(\cap_{r=1}^n A_r) \ge 1 - C_1 n e^{-\mu |\log \varepsilon| n^{\tau}/2}.
\end{equation}
The latter is essentially the event described in Lemma~\ref{lem:etaloc} when choosing $\varepsilon$ sufficiently small to absorb several constants (and allowing to choose any $\xi<\tau$). However, we need to show that all the $k_r$, $r=1,\ldots,n$, can be chosen distinct, so that an ONB of eigenvectors can be associated one-to-one with all localization centers in $[1,\ldots,n]$.

As in a similar context in \cite{EK}, a convenient way to do this is by using Hall's Marriage Theorem. For $U\subset [1,\ldots,n]$ let $\mathcal{N}(U) := \cup_{r\in U} \mathcal{N}_r$. We have to show that $|U| \le |\mathcal{N}(U)|$ for all $U$. By Hall's Marriage Theorem this is necessary and sufficient for the existence of a bijection $h:[1,\ldots,n] \to [1,\ldots,n]$ such that $h(r) \in \mathcal{N}_r$ for each $r$. Now the choice $k_r = h(r)$ yields the desired result.

Suppose, on the contrary, that $|U| \ge |\mathcal{N}(U)|+1$ for some $U$. If $j\notin \mathcal{N}(U)$ and $r\in U$, then $|\varphi_{E_r}(j)| < n^{-\alpha}$. Thus, for $r\in U$,
\begin{equation}
\sum_{k\in \mathcal{N}(U)} |\varphi_{E_r}(k)|^2 = 1- \sum_{j\notin \mathcal{N}(U)} |\varphi_{E_r}(j)|^2 > 1 - n^{1-2\alpha}.
\end{equation}
This implies
\begin{eqnarray}
|U|-1 \ge |\mathcal{N}(U)|  = \sum_{k\in \mathcal{N}(U)} \|\delta_k\|^2
& \ge & \sum_{k\in \mathcal{N}(U)} \sum_{r\in U} |\varphi_{E_r}(k)|^2 \nonumber \\ & > & |U| (1-n^{1-2\alpha}),
\end{eqnarray}
which gives $1\le |U| n^{1-2\alpha} \le n^{2-2\alpha}$, a contradiction to the choice $\alpha>1$.

\vspace{.5cm}

We now proceed with the proof of Theorem~\ref{thm:fock}. By Lemma~\ref{lem:etaloc}, with probability $1-e^{-n^{\xi}}$, there are normalized eigenfunctions $\varphi_k$ of $A$ satisfying (\ref{expdecay}). As $A$ is real, the $\varphi_k$ can be chosen real. They determine the orthogonal matrix $U$ in (\ref{diagA}) as $U(k,j) = \varphi_k(j)$.

To prove part (a) of Theorem~\ref{thm:fock} we expand the vectors $\tilde{\psi}$ in terms of the vectors $\tilde{e}$. This calculation is well known, for example it is not hard to see that the resulting (\ref{detformula}) below arises as a special case of (\ref{multi_cor}). We include the argument here for completeness.

From $\psi_0 = e_0$ and $\tilde{b}_k^* = \sum_{j=1}^n U(k,j) c_j^* = \sum_{j=1}^n \varphi_k(j) c_j^*$, we get
\begin{eqnarray}
\tilde{\psi}_{(k_1,\ldots,k_r)} & = & \prod_{m=1}^r \tilde{b}_{k_m}^* e_0 = \prod_{m=1}^r \left( \sum _{j_m=1}^n \varphi_{k_m}(j_m) c_{j_m}^* \right) e_0 \nonumber \\
& = & \sum_{j_1,\ldots,j_r \,\mbox{\tiny distinct}} \prod_{m=1}^r \varphi_{k_m}(j_m) c_{j_m}^* e_0 \nonumber  \\
& = & \sum_{j_1<\ldots<j_r} \sum_{\pi \in S_r} \left( \prod_{m=1}^r \varphi_{k_m}(j_{\pi(m)}) \right) \left( \prod_{m=1}^r c_{j_{\pi(m)}}^* \right) e_0 \nonumber \\
& = & \sum_{j_1<\ldots<j_r} \sum_{\pi \in S_r} \left( \prod_{m=1}^r \varphi_{k_m}(j_{\pi(m)}) \right) \times  \nonumber \\
& \mbox{ } & \quad \times (-1)^{sgn(\pi) + \sum_{m=1}^r j_m -r} \tilde{e}_{(j_1,\ldots,j_r)}.
\end{eqnarray}
Here we have used the CAR to normal order the $c_j^*$ and that, up to sign changes, $c_j^*$ acts on the spin basis by flipping up the $j$-th spin.
The final result for $\tilde{\psi}_{(k_1,\ldots,k_r)}$ becomes
\begin{equation}
 \sum_{j_1<\ldots<j_r} \left( (-1)^{\sum_{m=1}^r j_m - r} \det \left( \varphi_{k_m}(j_{\ell}) \right)_{m,\ell=1}^r \right) \tilde{e}_{(j_1,\ldots,j_r)}.
\end{equation}
From this we read off that for arbitrary $1\le r\le n$, $1\le k_1 < \ldots <k_r \le n$ and $1\le j_1 < \ldots < j_r\le n$,
\begin{equation} \label{detformula}
\left| \langle \tilde{\psi}_{(k_1,\ldots,k_r)}, \tilde{e}_{(j_1,\ldots,j_r)} \rangle \right| = \left| \det \left( \varphi_{k_m}(j_{\ell}) \right)_{m,\ell=1}^r \right|,
\end{equation}
while $\langle \tilde{\psi}_{(k_1,\ldots,k_r)}, \tilde{e}_{(j_1,\ldots,j_s)} \rangle=0$ if $r\not=s$ (which reflects particle number conservation).

Given exponential decay (\ref{expdecay}) of the one-particle Hamiltonian, we now get decay of the Slater determinant on the right of (\ref{detformula}) as a special case of a result from \cite{SimsWarzel}, stated as Theorem~\ref{thm:SW} in Section~\ref{sec: correlations} above:

Let $\omega$ be the $n\times n$-matrix with matrix elements $\langle \delta_j, \omega \delta_k \rangle = \varphi_k(j)$. This matrix is unitary and, in particular, $\|\omega\|=1$. To incorporate the result of \cite{EK} we choose the growth function $K$ piecewise as $K(\ell) = 0$ if $\ell < n^{\tau}$ and $K(\ell) = \ell$ if $\ell \ge n^{\tau}$. Then (\ref{expdecay}) says that $|\langle \delta_j, \omega \delta_k \rangle| \le e^{-\eta K(|j-k|)}$ with probability at least $1-e^{-n^{\xi}}$. Thus we can apply (\ref{det_dec}) for any $\eta_0<\eta$. For the given choice of $K$ we find $I(\mu_0) \le C(\mu_0,\tau) n^{2\tau}$. This readily yields (\ref{Fockloc}), completing the proof of part (a) of Theorem~\ref{thm:fock}.

The proof of part (b) is straightforward:  From (\ref{fockbasis}) and $c_x = \sum_{j=1}^n \varphi_j(x) \tilde{b}_j$ we get
\begin{eqnarray}
\langle \tilde{\psi}_k, n_x \tilde{\psi}_k \rangle & = & \langle \tilde{\psi}_k , c_x^* c_x \tilde{\psi}_k \rangle \\
\nonumber & = &\Scale[0.90]{\displaystyle \sum_{j_1,j_2} \varphi_{j_1}(x) \varphi_{j_2}(x) \langle \tilde{b}_{j_1} \prod_{m=1}^r \tilde{b}_{k_m}^* \psi_0, \tilde{b}_{j_2} \prod_{\ell=1}^r \tilde{b}_{k_{\ell}}^* \psi_0 \rangle} \\
\nonumber & = &\Scale[0.90]{\displaystyle \sum_{j_1} |\varphi_{j_1}(x)|^2 \| \tilde{b}_{j_1} \prod_{m=1}^r \tilde{b}_{k_m}^* \psi_0\|^2  =  \sum_{m=1}^r  |\varphi_{k_m}(x)|^2,}
\end{eqnarray}
where we have used that $\| \tilde{b}_j \prod_{m=1}^r \tilde{b}_{k_m}^* \psi_0\|^2$ is $1$ if $j \in \{k_1,\ldots,k_r\}$ and $0$ otherwise. By our assumption $\min_{\ell} \{|k_{\ell}-x|\} \ge n^{\tau}$ we can use (\ref{expdecay}) to bound this by
\begin{eqnarray}
\sum_{j = \min_{\ell} \{|k_{\ell}-x|\}}^{\infty} |\varphi_j(x)|^2 \le 2\sum_{j = \min_{\ell} \{|k_{\ell}-x|\} }^{\infty} e^{-2\eta j},
\end{eqnarray}
which is the bound claimed in (\ref{occnumberbound}).

%
%
%
%
%
%
%

\end{document}